\documentclass[times,twocolumn,final]{elsarticle} % here we use the article class, rather than elsarticle

\usepackage{cnf}
\usepackage{framed,multirow}

\usepackage{amssymb}
\usepackage{latexsym}

\usepackage{url}
\usepackage{xcolor}
\definecolor{newcolor}{rgb}{.8,.349,.1}

\usepackage{hyperref}

\usepackage[switch,pagewise]{lineno} %Required by command \linenumbers below

\journal{Combustion and Flame}

\usepackage{chemformula}

\begin{document}

\verso{Razavi et al.}

\begin{frontmatter}

\title{Pressure and temperature dependence of methane pyrolysis and hydrogen production in a micro flow reactor}

\author[1]{Mohammad Reza \snm{Razavi}\corref{cor1}}
\cortext[cor1]{Corresponding author: University of Toronto Institute for Aerospace Studies, 4925 Dufferin Street, Toronto, M3H 5T6, Canada.}
\emailauthor{m.razavi@mail.utoronto.ca}{Mohammad Reza Razavi}
%\ead{example@email.com}

\author[1]{\"{O}mer L. \snm{G\"{u}lder}}%\fnref{fn1}
%\fntext[fn1]{Author footnote 1.}  

\address[1]{University of Toronto Institute for Aerospace Studies, 4925 Dufferin Street, Toronto, M3H 5T6, Canada.}

\begin{abstract} % 100 to 300 words.
Methane pyrolysis experiments were performed in a micro flow reactor to investigate the chemical mechanism sensitivities at elevated pressures and temperatures. The micro flow reactor used in this study provides independent control over reaction parameters. Experiments were performed at temperatures from 1400 to 1800 K and pressures from 1 to 10 bar at a fixed residence time of 60 ms. The yields of hydrogen, unreacted methane, and solid carbon were quantified for comparison to predictive models. The mass of heterogeneous depositions was also quantified. The presence of \ch{C2} and PAH species were measured by GC-MS. The results were compared to 1D simulations using a discrete sectional model for capturing the kinetics of particulate formation and surface deposition. The simulations replicated the pressure trend well at high temperatures (1800 K), but the sensitivity to temperature was in disagreement. Simulations generally found lower carbon and hydrogen yields and a stronger sensitivity to temperature than the experiment. An abundance of PAH intermediates was measured at lower temperatures and higher pressures. The simulated surface deposition rates were lower than the measured values but showed similar profiles. The results demonstrate the utility of micro flow reactors in recreating pyrolysis environments for hydrogen production from methane. The predictive capabilities of simulations could be improved by advancing the fuel decomposition and PAH formation and consumption kinetics at low temperatures and by detailed treatment of surface site density.
\end{abstract}


\end{frontmatter}

%\linenumbers

\section*{Novelty and significance statement}

Pyrolysis studies in shock tubes or flow reactors typically cannot resolve overlapping processes governing carbon formation. A micro flow reactor is used to investigate carbon formation processes independently at intermediate residence times and high pressures — conditions for which experimental data remains limited. The study found opposite pressure dependencies between heterogeneous and homogeneous carbon formation rates and revealed their temperature sensitivity, trends that have not been previously resolved. PAH and \ch{H2} trends are reported which further enrich the modelling database. 

Current kinetic mechanisms for gas phase carbon formation and surface deposition predict higher onset temperatures and lower deposition rates than measured and disagree with experimental pressure dependencies and temperature sensitivities for some conditions. Insufficient surface reactions, constant site density, and fuel decomposition reactions are identified as key areas for improvement. Models resolving surface deposition and gas phase trends and intermediates at wider conditions can improve predictive capabilities for pyrolysis reactor design.

\section{Introduction}
\label{sec:introduction}

Methane pyrolysis can be an effective substitute for conventional steam methane reforming for \ch{H2} production, retaining the carbon content in solid phase. It has benefits of both preventing \ch{CO2} emissions and providing a source for carbon black production \cite{methPyrolReview}. However methane pyrolysis reactions can be difficult to characterize. The enhancement of commercial methane pyrolysis processes relies on accurate kinetics and requires better understanding of the dependencies of carbon formation mechanisms \cite{methPryolReviewTech}.

For commercial hydrogen production, these dependencies are studied at low temperatures (less than 1300 K) and often in the presence of catalysts. However in combustion environments such as in gas turbines or diesel engines, fuel pyrolysis contributes to formation of soot particulates at flame temperatures (about 2200 K) and pressures up to 250 bar \cite{GasTurbineRT, dieselPressure}. Investigating methane pyrolysis at engine relevant conditions can provide a more robust understanding of the underlying kinetics and lead to better characterization of the carbon formation processes.

Soot formation in non-premixed methane combustion follows mainly the chemistry of methane pyrolysis. This begins with fuel decomposition and polycyclic aromatic hydrocarbon (PAH) formation. Their mechanisms have been developed extensively based on shock tube and flow reactor measurements \cite{BrezinskyPAHChem_RL, CreckDev_RL}. Understanding of nucleation and soot growth has also been developed from shock tubes \cite{WagnerSTPyro_RL}, laminar diffusion flames (LDF) \cite{gigoneTEM} and turbulent flames \cite{OLGTurbulent}, counterflow diffusion flames (CDF) \cite{YaleSootPres}, and flow reactors \cite{methanePFR}, among others. Discrete sectional models combining the soot growth kinetics with fuel and PAH chemistry work well to cover a range of reaction domains \cite{Cuoci_pyrolysis_benzene}. However, further investigation is needed to thoroughly characterize the soot and carbon formation processes and how they vary under different operating conditions. The complicated kinetics of fuel decomposition, nucleation, and gas-solid chemistry makes this characterization challenging \cite{olgSootReview}.

\begin{figure}
	\centering
	\includegraphics[width = 192pt]{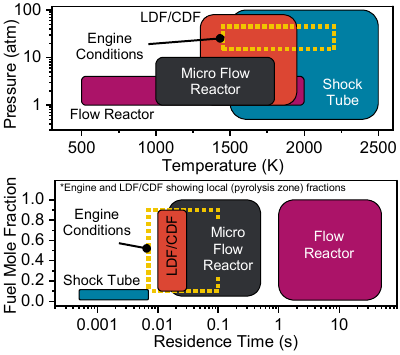}
	\caption{Typical reactor operating regimes for methane pyrolysis studies in shock tubes, flow reactors, and micro flow reactors. Soot forming combustion conditions in non-premixed LDFs, CDFs, and engines shown assuming local mole fractions in the pyrolytic flame zone. For surveyed studies see the supplementary material.}
	\label{fig:reactionregimes}
\end{figure}

Studies in both shock tubes and flow reactors can address this challenge but with limitations. Although shock tubes can provide high temperatures and pressures and are used to study methane pyrolysis \cite{SchulzST, Ferris2024}, the isothermal condition fails for high fuel fractions undergoing endothermic pyrolysis. Flow reactors are more restrained in the reaction conditions, but offer more flexibility with the fuel composition and operate over longer residence times \cite{FlowRMethane,FlowRMethaneHT}. Micro flow reactors providing isothermal conditions for high fuel fractions are an alternative method, enabling isolation of reaction parameters and flexibility in fuel composition with access to a wide range of residence times. They have been adopted both for PAH chemistry and soot formation studies \cite{marutaGCMS, Maruta2022}. Typical operating conditions from a survey of several shock tube and flow reactor experiments are shown in Fig. \ref{fig:reactionregimes} along with the operating conditions of the micro flow reactor used in the current work. While some overlap exists in the pressure-temperature regimes between the three reactor types, the micro flow reactor provides access to novel mass fractions and residence times in between shock tube and flow reactor conditions.

\subsection{Soot formation in gas phase}

For gas turbines typical residence times in the combustor range from 10 to 20 ms (\cite{GasTurbineRT}) and diesel engines can vary around 100 ms. During this time many combustion processes take place in addition to gas phase pyrolytic soot formation such as spray vapourization and oxidation. Due to the highly turbulent conditions the time history of the pyrolysis reactions becomes intractable. In laboratory LDF and laminar CDF this temperature time history is better characterized but not constant, with typical residence times of several tens of ms over temperatures varying from 1500 K to 2000 K \cite{olgSootReview}. The typical operating pressures, temperatures, residence times, and fuel mass fractions for gas phase soot formation in both engines and laboratory flames are also shown in Fig. \ref{fig:reactionregimes}. The micro flow reactor can access the engine relevant gas phase soot forming conditions under constant temperature time history.

In this work a micro flow reactor was used to study methane pyrolysis at high temperature, high pressures, and residence times typical of gas phase soot formation at engine relevant conditions. Yields of major gas species and both heterogeneous and homogeneous carbon were quantified. Experimental findings were compared to numerical simulation results obtained from the latest sectional models of soot formation.

\section{Methodology}
\label{sec:exp}
\subsection{Experimental}

\begin{figure}[t!]
\centering
%\vspace{-0.4 in}
\includegraphics[width=192pt]{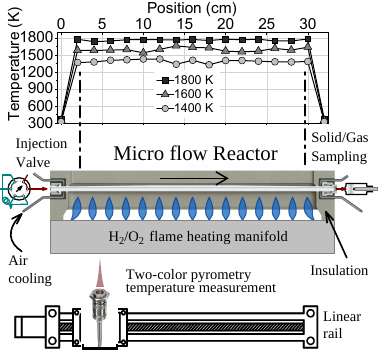}%
\vspace{2 pt}%
\caption{Schematic of the micro flow reactor.} 
\label{Fig:exp}
\end{figure}

A micro flow reactor was built to provide isothermal high pressure pyrolysis conditions. The main components consist of a ceramic tube (3.175 mm ID) where the reaction takes place, a heated 6-way valve (Valco) which prepares the inlet mixture, and a solid and gas sampling system for measurement of the products. A schematic of the apparatus is shown in Fig. \ref{Fig:exp}. Reactor heating is provided by \ch{H2}/\ch{O2} flames, and temperature measurement by two-color pyrometry (Optris CTRatio). Mixtures of \ch{CH4} diluted in \ch{Ar} were preheated in the sample loop of the 6-way valve and sent through the reactor using \ch{N2} carrier gas. The novelty of the apparatus is in combining an injection valve with a plug flow reactor, thereby providing control of the temperature, composition, and critically the volume of the injected fuel mixture. This enables intermittent reactor operation for durations between 1 to 5 seconds while keeping reaction residence time constant between 15 to 100 ms. The controlled duration limits the accumulated carbon mass and enables carbon yields to be quantified. 

At the reactor exit, gas products were sampled into evacuated volumes of a sample trapping valve (Valco) through a heated probe while solid products were collected on a porous alumina filter. Solid products deposited on the tube walls were collected by cutting the reactor into 20 mm segments along the length. The cut segments and the porous filter were washed with deionized water and dried in a thermal vacuum prior to carbon burnoff measurement (Eltra SC800) for quantification of total solid carbon mass. For additional details the reader is refereed to the supplementary material. Collected gas samples were split and measured through GC-MS for qualitative analysis of PAH and GC-TCD for quantification of light gas species. Multiple gas samples were collected per run by repeating the fuel mixture injections and purging with \ch{O2} between each injection. The collected gas samples were analysed through an automated injection sequence. 

\subsection{Isothermal reaction conditions}

Evaluation of the isothermal assumption in the micro flow reactor was performed by simulations, internal temperature measurements, and experiments. Simulations using 2D non reacting flow with heat transfer were performed to assess heat transfer rates and internal temperature profiles. The analysis found the carrier gas temperature at the centerline reaches the wall temperature within 1 to 3 ms. 

Internal temperature measurements were taken using a 1.6 mm diameter C-type thermocouple positioned inside the reactor. The closed radiative environment prevents any heat loss in the thermocouple and thereby the measured values are representative of the internal wall temperature. The thermocouple was used to calibrate the two-color pyrometry instrument to represent internal wall temperatures by observation of the external wall temperature. The temperature drop was then observed during pyrolysis reactions and found temperature drops by a maximum of about 100 K for the highest carbon loading cases. This is within the temporal and spatial fluctuations of wall temperature on the order of $\pm$ 50 K. Typical wall temperature profiles are shown in the inset in Fig. \ref{Fig:exp}. Measurement uncertainty analysis is provided in Section \ref{sec:UC}.

\begin{figure}[htpb!]
	\centering
	%\vspace{-0.4 in}
	\includegraphics[width=192pt]{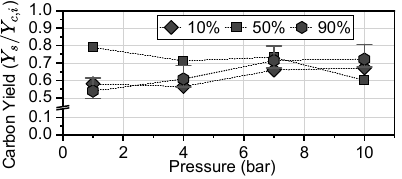}%
	\caption{Carbon yields from pyrolysis of methane at 10, 50, and 90\% fuel mass fraction, 60 ms residence time, 1800 K.} 
	\label{Fig:IsoMF}
\end{figure}

Further experiments were performed with varying mass fractions of 10, 50, and 90\% methane in Ar at 1800 K, from 1 to 10 bar, Fig. \ref{Fig:IsoMF}. For both the 10\% and 90\% mass fractions the carbon yield increased with increasing pressure, and had similar absolute values. For the 50\% fuel mass fraction condition the carbon yields decreased with increasing pressure, but closely matched the 10 and 90\% conditions in absolute values. The differing trends are attributed to shifts in underlying mechanisms sensitive to mass fraction. They are not attributed to temperature shifts due to endothermic pyrolysis, as all mass fractions (particularly 10 and 90\%) found similar yields. Despite the 90\% case having 9 times higher energy loss from pyrolysis compared to the 10\% case the yields remained comparable, indicating the heat transfer rate is far higher than the heat absorption rate and the isothermal assumption is valid. To facilitate the instrument limits of carbon deposition measurements simultaneously with gas species measurements, the 50\% mass fraction condition was selected for subsequent experiments.

\subsection{Reaction onset temperature}

\begin{figure}[t]
	\centering
	%\vspace{-0.4 in}
	\includegraphics[width=192pt]{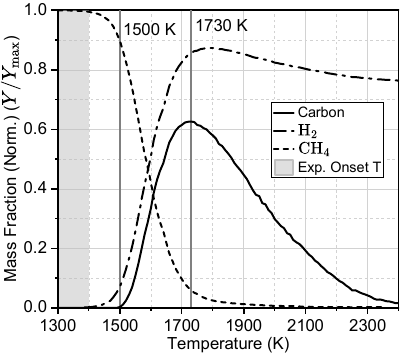}%
	\caption{Simulation results of methane pyrolysis at 60 ms residence time and 1 atm pressure using the CRECK mechanism. Simulated and experimental soot onset temperatures shown.}
	\label{Fig:OnsetT}
\end{figure}

Additional simulations of the pyrolysis reaction were performed across a range of reaction temperatures, see Fig. \ref{Fig:OnsetT}. Details of the simulation mechanism are provided in Section \ref{sec:ExpSim}. These were fixed at a residence time of 60 ms and a pressure of 1 atm. The simulations found an onset temperature for carbon formation of approximately 1500 K and maximum carbon mass fractions at 1730 K. The range of temperatures selected for the experiment covers this domain from 1400 K to 1800 K. Experimentally it was found that at 1300 K, there was no carbon formation for the 60 ms residence time, while significant yields were measured at 1400 K. This suggests a true onset temperature between 1300 and 1400 K. 
 
This range is in agreement with measurements in LDF and spray pyrolysis experiments \cite{SaitoLDF, GomezLDF, SkeenSpray_RL}. For \ch{CH4} Saito et al. found a soot onset temperature of 1390 K along the centerline of LDFs \cite{SaitoLDF}. For other fuels up to benzene the onset temperature varied between 1300 and 1400 K \cite{SaitoLDF, GomezLDF}, while for \textit{n}-dodecane spray pyrolysis it was measured at 1450 K \cite{SkeenSpray_RL}. In LDFs the onset temperature is reduced by the presence of trace oxidative radicals which are not present in the current study. Koshi et al. suggested that the surface depositions in high surface area flow reactors may have a disordered structure acting as a catalytic pyrocarbon \cite{MethProblems}. The catalytic effect of surface depositions may substitute the effect of trace radicals found in LDFs and lead to the good agreement observed in onset temperature. The selected experiment temperature range of 1400 to 1800 K captures both onset carbon formation and high temperature phenomenon.

\subsection{Simulations and kinetic mechanisms}
\label{sec:ExpSim}

Simulations of a 1D isothermal plug flow reactor (PFR) were performed using the OpenSMOKE++ \cite{opensmoke} solver. The CRECK mechanism \cite{Polimi2019} optimized for \ch{CH4} (C1-C16 + Soot) was used to simulate soot formation kinetics including a gas phase sectional model for tracking soot growth. The model has been extensively validated for many flame types (see for example \cite{Cuoci_pyrolysis_benzene}), however further work remains to address the complexities of pyrolysis reactions \cite{MethProblems}.

Simulations were also performed using a surface kinetics mechanism initially developed for vapour deposition \cite{SurfMechOG} but modified with improved reaction rates and combined with the CRECK mechanism for gas phase kinetics \cite{Cuoci_SurfModel}. The mechanism is validated with pyrolysis experiments in perfectly stirred reactors (PSR) and PFRs for a range of conditions \cite{Cuoci_SurfModel, SurfMechOG}. Simulations using the combined mechanism here referred to as CRECK-Surf were performed with a series of 1D PSRs in the OpenSmoke++ framework \cite{opensmoke}. Due to the stiffness of the kinetic mechanism at high temperatures, direct PFR solutions were not obtainable. Surfaces in each PSR were initialized using equal distributions of zig-zag and armchair sites as suggested in \cite{Cuoci_SurfModel} and \cite{SurfMechOG}. Surface site density was kept constant at $1.33\mathrm{x}10^{-9}$ mol/$\mathrm{cm}^2$. To replicate equivalent PFR results, 15 PSR simulations in total were linked for each condition and adjusted to match the surface area of the 15 cut segments in the micro flow reactor experiment. Calculated surface deposition rates were scaled by the experiment duration (between 100 to 1000 ms) to obtain predictions of deposition mass for comparison to experimental measurements.

\subsection{Uncertainty analysis}
\label{sec:UC}

A C-type thermocouple (5\% W, 26\% Re) was used for internal wall temperature measurements. Between 1400 and 1800 K, it has maximum reported deviation of 5 K from ASTM E-230 standard curve. The thermocouple was used for a 5-point calibration of the emissivity value for the two-color pyrometer between 1300 and 1900 K. Each calibration point temperature was held for several minutes to establish steady state conditions between the inner wall temperature and the thermocouple. Then, the temperature reading was averaged over 2 minutes to obtain the calibration point. The calibration points were used to adjust the assumed emissivity to match the external wall temperature measurement to the thermocouple reading. The standard accuracy of the pyrometer is $\pm$ (0.5 \% of reading $+$ 2 $^\circ $C) and the repeatability is $\pm$ 0.3 \% of reading. Combined uncertainties for the thermocouple, 5-point calibration, and pyrometer give a temperature measurement uncertainty of $\pm$ 10 K. 

\begin{figure}[h]
	\centering
	\includegraphics[width=192pt]{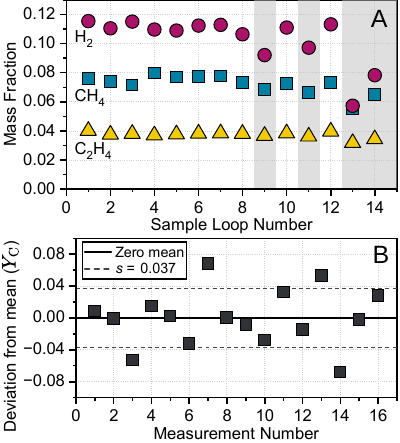}
	\caption{\textbf{A:} Representative gas species measurements for a single experiment run at 1600 K, 7 atm, using the gas sampling system and GC-TCD. Measurements show the variance between each sampling loop. Samples rejected due to leakage shown in grey. \textbf{B:} Deviation from the mean of carbon yield measurements from multiple experiment runs at varying conditions. Mean values shifted to zero for all conditions. Standard deviation on carbon yield measurements found as s = 0.037.}
	\label{Fig:GasREP}
\end{figure}

For gas species measurement 5 to 14 samples were collected consecutively in each experimental run. Representative measurements of collected samples are shown in Fig. \ref{Fig:GasREP}a. Leaks between loops are minimal and neglected as all samples are repetitions. Leaks out of some loops are apparent in high diffusivity species such as hydrogen, seen in loop 9, 11, 13, and 14 in Fig. \ref{Fig:GasREP}a. All the gas species measurements from these loops are rejected from the final reporting including heavier species which show no leakage. For PAH measurements, the first few samples are rejected due to skewed readings owing to adsorption along the sampling path. Once saturated, subsequent loops found more consistent measurements. The accepted measurements are averaged to obtain gas species mass fractions at one condition. This process is repeated 1 to 3 times for each condition. Single run repeatability (between different loops) is well within the uncertainty of the GC-TCD measurement. Inter run repeatability (measurements from independent experimental runs) can show variations of about 5 \%, owing to variability in the reactor temperature and the thermal conditions of the gas sampling probe and sampling lines. The reported measurements are averages of between 5 to 10 loops and 1 to 3 repetitions of each condition, and show an average variation of 7 \%. The estimated instrument uncertainty including the GC-TCD calibration mixtures and detector response is 5 \%. The combined standard uncertainty for each measured gas mass fraction is 9 \%.

For carbon yield, experiments were repeated 1 to 3 times for each condition and the measured yields were averaged. The yields were corrected for baseline carbon determined through blank measurements. The deviation of yield from the zero centered mean at each condition is shown in Figure \ref{Fig:GasREP}b. The standard deviation is s = 0.037, and is mostly attributed to variations in reactor temperature between each repetition. The blank value varies between 60 to 160 $\mu$g C. The inlet carbon mass fraction has $\pm$ 1 \% error from the mass flow controllers supplying the \ch{CH4} and \ch{Ar}. Finally up to 5 \% of each cut segment from the reactor walls may be lost due to the blade thickness. Combining these error sources gives the uncertainty for carbon yield as $+$ (6 \% + 0.042), $-$ (1 \% + 0.037).

\section{Results and Discussion}
\label{sec:results}

\begin{figure}[!ht]
	\centering
	\includegraphics[width=192pt]{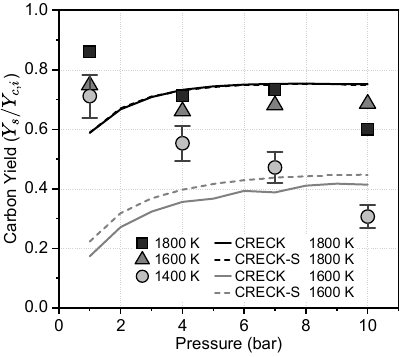}
	\caption{Carbon yield from pyrolysis of \ch{CH4}, 60 ms residence time, 50\% mass fraction in \ch{Ar}. Simulations with CRECK and CRECK-Surf mechanism. Error bars are typical.}
\label{fig:carb}
\end{figure}

Methane pyrolysis reactions were studied at a range of conditions (see Table \ref{Tab:expCond}). The selected range accommodates sufficient gas product volumes for GC-MS and GC-TCD detection, while limiting solid product mass within the limits of carbon burnoff measurement. The residence time was kept constant at 60 ms for all conditions. Measurements of total solid carbon mass were normalized by the carbon mass in the injection volume to give the carbon yield, $CY$, as

\begin{equation}
CY = m_\mathrm{c}/(Y_\mathrm{\ch{C}} V_\mathrm{i} \rho_{\mathrm{i}}) = Y_{\mathrm{s}}/Y_{\mathrm{c,i}} , 
\end{equation}
where $m_{\mathrm{c}}$ is the measured carbon mass, $Y_{\ch{C}}$ the carbon mass fraction in the inlet mixture, and $V_\mathrm{i} \rho_{\mathrm{i}}$ the total mass of the injected volume. The solid mass fraction ($Y_{\mathrm{s}}$) and carbon mass fraction at the inlet ($Y_{\mathrm{c,i}}$) can be similarly used when available from simulation results. Measured and simulated carbon yields from methane pyrolysis are shown in Fig. \ref{fig:carb}. 

\begin{table}[!ht]
	\footnotesize%
	\caption{Range of experimental conditions}
	\label{Tab:expCond}
	\centerline{\begin{tabular}{lcl}
			\hline 
			Parameter							&	Value				& Units	\\
			\hline
			Fuel mass fraction in \ch{Ar} 		&	50				& \%	\\
			Temperature 						&	1400, 1600, 1800	& K		\\
			Pressure 							&	1, 4, 7, 10			& bar	\\ 
			Residence time 						&	60					& ms	\\
			\hline 
	\end{tabular}}
\end{table}

\begin{figure}[!t]
	\centering
	\includegraphics[width=192pt]{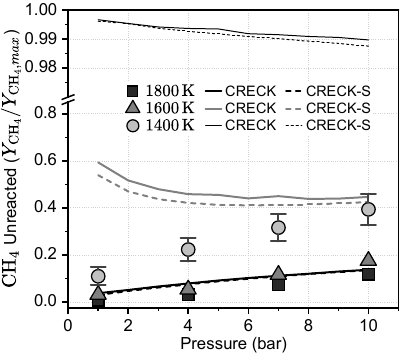}
	\caption{Normalized fraction of unreacted \ch{CH4} from pyrolysis reactions, 60 ms residence time, 50\% mass fraction in \ch{Ar}. Simulations with CRECK and CRECK-Surf mechanism. Error bars are typical.}
	\label{fig:meth}
\end{figure}

\begin{figure}[!t]
	\centering
	\includegraphics[width=192pt]{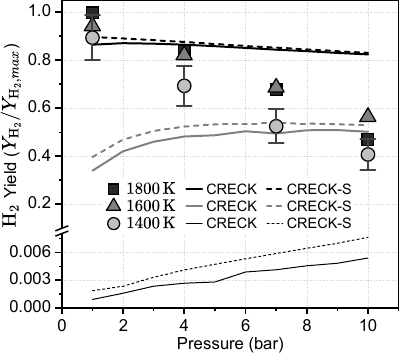}
	\caption{\ch{H2} yield from pyrolysis of \ch{CH4}, 60 ms residence time, 50\% mass fraction in \ch{Ar}. Simulations with CRECK and CRECK-Surf mechanism. Error bars are typical.}
	\label{fig:hydr}
\end{figure}

Major gas species in the products are unreacted \ch{CH4} and \ch{H2}. The \ch{Ar} mass fraction in the injection mixture was used as an internal standard to calculate the mass fraction of products independent of the level of dilution with \ch{N2}, which was used both as the reactor carrier gas and for sample gas transfer. The measured and simulated normalized mass fractions of unreacted \ch{CH4} and \ch{H2} are shown in Figs. \ref{fig:meth} and \ref{fig:hydr}, respectively. The measured values correspond to the time-averaged composition over the entire experimental duration, as the sampled volume is comparable to the injected volume. Considering the short residence time of 60 ms and the experimental duration of about 1 second, the measured composition is near to the steady state value and is comparable to simulation results.

\subsection{Reaction mass balance}

\begin{figure}[h]
	\centering
	\includegraphics[width=192pt]{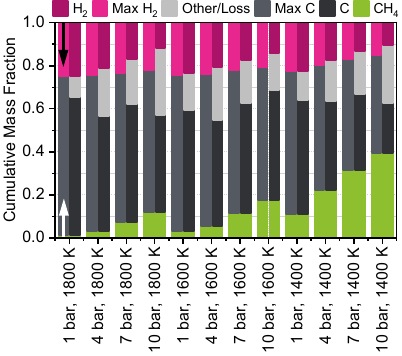}
	\caption{Cumulative mass fraction of reaction products. Average mass recovery for quantified species was 80 \% ($+$ 7 \%, $-$ 5 \%).}
	\label{Fig:cumulMF}
\end{figure}

The total mass balance of the reaction is plotted as a cumulative mass fraction of the measured species, shown in Fig. \ref{Fig:cumulMF}. Measured but unquantified species include \ch{C2} and PAH up to \ch{C12}. These are estimated to account for 5 to 10 \% of the mass fraction and are included as "other" species, Fig. \ref{Fig:cumulMF}. The maximum possible carbon and \ch{H2} mass fractions are calculated based on complete conversion of the consumed methane. The difference between the measured carbon and maximum carbon indicates how much of the carbon mass remains in the intermediates or is lost in the measurement process. The same applies to \ch{H2}. Across all experiment conditions the average mass recovery was 80\% ($+$ 7 \%, $-$ 5 \%) for calibrated species and about 90\% including detected but unquantified species.

\subsection{Pressure and temperature dependence}
\label{subsec:resultsPressure}

The carbon yield was found to decrease with increasing pressure across the range of temperatures investigated, Fig. \ref{fig:carb}. The trend of pressure dependence was evaluated by fitting the data to a power law where the exponent of pressure is $n = \mathrm{log}(Y/Y_{\mathrm{atm}})/\mathrm{log}(P/P_{\mathrm{atm}})$. The 1400 and 1800 K conditions show a dependence of $n = -0.28$ and $n = -0.13$ respectively, while the 1600 K condition shows a stronger pressure dependence with $n = -0.05$. The reduction of carbon yield with pressure is consistent with other flow reactor studies, which attribute the trend to the suppression of chain branching reactions governing \ch{CH4} decomposition \cite{FlowRMethane}. However, the simulated carbon yield increases with increasing pressure at all temperatures. At 1800 K the predicted carbon yield values are in agreement with the experiment measurements, but at 1600 and 1400 K they are lower than the measurements. This strong sensitivity of carbon yield to temperature was not observed in the experiments. The simulations display higher pressure sensitivity at 1600 K, similar to the experiment findings.

Carbon yield decreasing with pressure is in stark contrast with LDFs, which find maximum soot yield scaling with $P^n$ where $n \simeq 2$ \cite{OLGLDF}. While carbon formation from pyrolysis and soot formation from combustion follow similar mechanisms, the absence of oxidative species in pyrolysis may limit the available pathways for radical formation and surface growth. In numerical simulations of soot formation in a LDF, the soot formation pressure dependence shows sensitivity to both oxidative radicals and surface growth mechanisms \cite{OLG_sim_LDF}. Normalized soot production rates in CDFs also find strong pressure dependencies of about $n \simeq 1.7$ \cite{SawanniPCI}. Numerical analysis of the formation processes in CDFs suggests the pressure dependence has strong sensitivity to the precursor composition, which itself depends on temperature variations \cite{SawanniCNF}. The absence of oxidative species and the constant reaction temperature are key differences between PFR and LDF/CDF which may explain the observed discrepancy in carbon and soot yield, however further investigation is needed to address this disagreement.

The unreacted methane fraction was found to increase with increasing pressure, Fig. \ref{fig:meth}. Like the carbon yield, this trend is also attributed to suppression of the chain branching reactions by pressure \cite{FlowRMethane}, which for the low residence time considered in this study would result in lower fuel consumption and thus higher fractions of unreacted \ch{CH4} remaining. The decrease in the methane decomposition and suppression of the chain branching reactions may subsequently inhibit radical formation and growth of intermediate species involved in \ch{H2} formation. This may explain the reduction of \ch{H2} yields with increasing pressure observed in the current study, Fig. \ref{fig:hydr}.  

The pressure dependence of unreacted \ch{CH4} is predicted well at 1800 K by all the simulations. However at lower temperatures they show reducing \ch{CH4} fractions with increasing pressure, Fig. \ref{fig:meth}. The reduction in \ch{H2} yield with increasing pressure is also captured well at 1800 K, but at 1600 and 1400 K the simulated \ch{H2} yield shows increasing yields with pressure, Fig. \ref{fig:hydr}. The pressure dependence order of \ch{H2} yield and unreacted \ch{CH4} was found to be similar at all the temperature conditions. The order was averaged between the temperatures and was found to be $n = -0.25$ and $n = 0.78$ for \ch{H2} yield and unreacted \ch{CH4}, respectively. The simulations predict strong temperature sensitivity of carbon yield, \ch{H2} yield, and unreacted \ch{CH4} fraction at the pressures investigated. They predict a significant reduction in reactant conversion at 1600 K while the experiments show a slight reduction beginning at 1400 K. This discrepancy in the onset temperature may be explained by low global reaction rates at the lower temperatures.

\subsection{Heterogeneous pyrolysis effects}

\begin{figure}[!t]
\centering
\includegraphics[width=192pt]{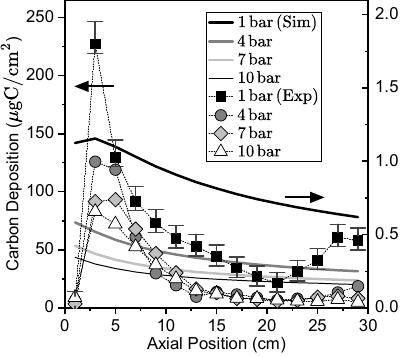}
\caption{Axial profiles of solid carbon deposition in micro flow reactor at 1600 K and varying pressures. Reactor inlet is at 0 cm. Experimental measurements shown (left axis). Error bars are typical. Simulated profiles using the CRECK-Surf mechanism with chained PSR shown (right axis).}
\label{fig:profiles_T}
\end{figure}

\begin{figure}[!t]
	\centering
	\includegraphics[width=192pt]{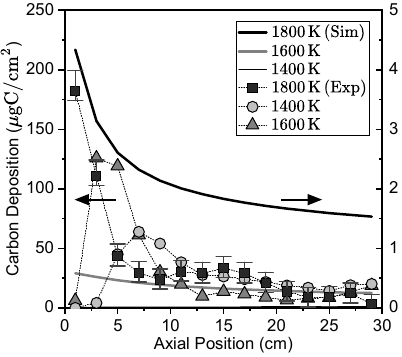}
	\caption{Axial profiles of solid carbon deposition in micro flow reactor at 4 bar and varying temperatures. Reactor inlet is at 0 cm. Experimental measurements shown (left axis). Error bars are typical. Simulated profiles using the CRECK-Surf mechanism with chained PSR shown (right axis).}
	\label{fig:profiles_P}
\end{figure}

\begin{figure}[!ht]
	\centering
	\includegraphics[width=192pt]{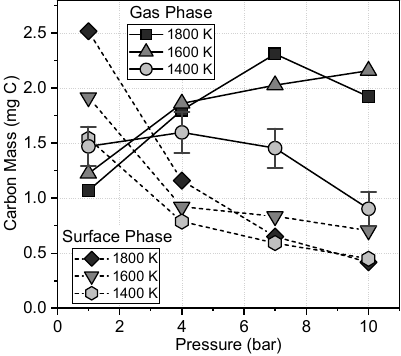}	
	\caption{Total gas phase (homogeneous) and surface phase (heterogeneous) carbon measurements from \ch{CH4} pyrolysis at 60 ms residence time. Sum of values corresponds to global carbon mass. Values are corrected to 4.2 $\mathrm{mg}$ injected carbon for consistency between conditions.}
	\label{fig:CMass_Combined}
\end{figure}

Representative carbon burnoff measurements of surface deposition profiles in the micro flow reactor found significant deposition mass along the reactor length, shown for varying pressures at 1600 K in Fig. \ref{fig:profiles_T} and for varying temperatures at 4 bar in Fig. \ref{fig:profiles_P}. The measurements are discrete measurements of 20 mm length segments of the reactor and approximate the true deposition profile. The highest deposition mass for most conditions was found near the reactor inlet and decreased steadily towards the center of the reactor. A slight increase in deposition mass was observed near the exit for some conditions (see Fig. \ref{fig:profiles_T} at 1 bar). This effect may be partially due to deviations from the nominal temperature in the temperature profile, which decreased slightly near the ends of the reactor and may induce a thermophoretic force on the mobile carbon particles. It may also be attributable to the kinetics of a secondary reaction regime observed in some shock tube pyrolysis experiments \cite{EreminSTPyro_RL, MathieuSTSoot_RL}. Increasing reaction pressure was found to suppress the deposition mass across the reactor length, while the position of the maximum deposition mass did not shift significantly. Gas phase carbon and surface deposition measurements are shown separately in Fig. \ref{fig:CMass_Combined}. The suppression of surface deposition by increasing pressure is consistent across all temperatures, and may be due to decreased diffusivity of the gas phase carbon at higher pressures. This is also suggested through modelling studies which found particle diffusion to significantly affect surface deposition rates \cite{JuanEavesThomson_2D_Diffusion}.

Simulation results using the CRECK-Surf mechanism show similar trends to the experimental measurements, Fig. \ref{fig:profiles_T}. The simulations predict high deposition masses at the reactor inlet which steadily decrease along the reactor length as the reaction progresses. They also replicate the suppression of deposition mass with increasing pressure, showing the strongest suppression between 1 and 4 bar in agreement with the experimental findings. However the simulated deposition mass is lower than the measured mass by a factor of about 100 for the range of pressures at 1600 K.

Increasing temperatures from 1400 to 1800 K increased the deposition mass and shifted the position of maximum deposition. At 1800 K, maximum deposition mass was found at the reactor inlet while at 1400 K it was found at about 7 cm past the reactor inlet and was lower. This reduction in deposition mass and shift in the peak location was observed despite the global yields at 4 bar remaining about constant with decreasing temperature. Furthermore at 1 bar the surface phase carbon deposition mass decreased with decreasing temperatures while the gas phase carbon increased. Since these trends are opposite and not consistent across all pressures, the reduction in surface deposition cannot be explained solely by gas phase kinetics and reflects the sensitivity of the surface deposition rates to temperature.

Simulations of the deposition profile capture the initial peak well at 1800 K, shown in Fig. \ref{fig:profiles_T}. However for 1600 and 1400 K, they predict no shift in peak location. The simulated surface deposition predicts stronger sensitivity to temperature than pressure, predicting a significant reduction with reducing temperature. This is similar to the results of the gas phase kinetics models which show low reaction yields at 1400 K. This finding may indicate that the gas phase reactions in the CRECK-Surf mechanism are not significantly influenced by the surface kinetics at the current conditions. This is also in agreement with the experimental measurements which found that comparable carbon yields at different temperatures can have varying amounts of surface carbon deposition, Fig. \ref{fig:CMass_Combined}. These findings suggest the importance of heterogeneous pyrolysis effects on the global yield may be challenged in the micro flow reactor experiment particularly at high pressures where the relative contribution of surface carbon to global yields is reduced.

The simulated deposition mass at 4 bar is lower than the experiment measurements by a factor of about 50 and 100 at 1800 K and 1600 K respectively. The lower surface deposition predicted by the simulations may be due to the residence time of the experiments in this study being 60 ms, significantly shorter compared to the validation experiments considered in \cite{Cuoci_SurfModel} which used residence times from 0.5 to 4 s. A heuristic assessment of the surface reactions in the CRECK-Surf mechanism found the majority involve reactions with short chain \ch{C2} to \ch{C4} species and some aromatics. The low residence times may limit formation of these key species in the simulations. Furthermore, the inclusion of heavier PAH may be key in capturing the deposition phenomenon. Detailed microscopic investigation of the pyrolytic deposits in microwave formed carbon identified the depositions to be a mixture of graphitic and amorphous segments with a liquid-like film originating from PAH \cite{ThomsonDetailedMicroscopy}. This composition provides exposed active sites and is suggested to catalyze pyrolysis \cite{MethProblems}. This would both reduce the onset temperature and accelerate global reaction rates compared to the gas phase kinetics, possibly explaining the high surface depositions measured in the current work and the temperature sensitivity found in the simulations which do not consider the development of the depositing surface. Analysis of the relevant \ch{C2} to \ch{C4} and PAH species could provide additional insight to address the discrepancy.

Investigations in a packed flow reactor at 1400 K and atmospheric pressure found the proportion of deposited carbon is increased at higher surface-to-volume (S/V) ratios \cite{SVRatioItalians}. They also found morphological differences with heavy deposition owing to higher S/V, however noted that temperature plays a more critical role. High temperatures resulted in less amorphous carbon and more graphitic carbon \cite{SVRatioItalians}. Although the phenomenon governing the proportion of deposition mass, its morphology, development of active surface sites, and subsequent influence on the gas phase kinetics are difficult to isolate in such experiments, the complex interplay remains an area for investigation. Understanding the relative significance of S/V ratio and residence times can provide useful insights into the sensitivity of the observed trends to these surface effects.

\subsubsection{Extended S/V ratios and residence times}

\begin{figure*}[ht!]
	\centering
	\includegraphics[width=\textwidth]{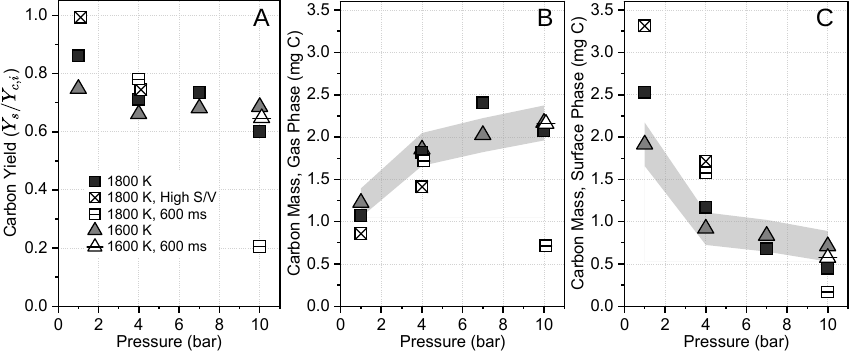}
	\caption{Carbon yields (A), gas phase carbon (B), and surface phase carbon (C) measurements for conditions with increased S/V and residence time. Also shown is corresponding nominal conditions. Error shown as shading for clarity.}
	\label{fig:ExtraConditions}
\end{figure*}

The primary experiments in the current study (60 ms residence time, S/V = 12.6 $\mathrm{cm}^{-1}$) were repeated at key conditions with a reduced diameter reactor (S/V = 25.2 $\mathrm{cm}^{-1}$) or increased residence time (600 ms). The experiments with higher S/V ratio were performed at 1800 K while experiments with higher residence time were performed at both 1800 K and 1600 K. Carbon yield measurements for the additional conditions are shown overlaid with the corresponding primary conditions in Fig. \ref{fig:ExtraConditions}. Separated measurements of gas phase and surface phase carbon mass are also shown. 

At 1800 K and 4 bar increasing the residence time increased the carbon yield owing to an increase in surface phase carbon while the gas phase carbon remained unchanged. For the 10 bar condition increasing the residence time at 1600 K had no discernible impact on the carbon yields, while at 1800 K higher residence times found much lower carbon yields. This result remains unexplained and suggests an apparent reverse mechanism for pyrolysis and deposition at longer residence times and high temperatures. These findings also suggest that the equilibrium condition is not readily reached for the 60 ms residence time and further gas and surface reactions are ongoing. The observed trends and sensitivities are thus representative of carbon formation and deposition kinetics.

Increasing the S/V ratio at 1800 K and 1 and 4 bar increased the surface phase carbon mass and reduced the gas phase carbon mass. As a result the total carbon yields increased by about 16 and 7 \% for 1 and 4 bar respectively. The proportion of surface phase carbon also increased from 70 \% to 79 \% at 1 bar and from 39 \% to 55 \% at 4 bar. Considering that the S/V of the reactor was doubled, the increase in the proportion of surface deposition was mild. Furthermore the pressure dependence of both the gas phase and surface phase measurements are unchanged. As such the increase in reactor S/V at the conditions in this study had a weak effect on the yields of heterogeneous and homogeneous carbon. Detailed investigation of its effect on the deposition morphology remains a subject of future work.

\subsection{PAH Measurements}

\begin{figure}[!t]
\centering
\includegraphics[width=192pt]{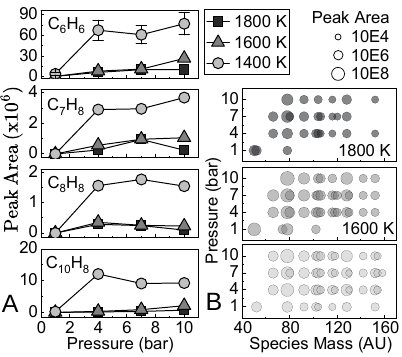}
\caption{GC-MS measurements of PAH from \ch{CH4} pyrolysis in a micro flow reactor. \textbf{A:} Pressure and temperature dependence of average peak area in the 4 most abundant species detected (\ch{C6H6}, \ch{C7H8}, \ch{C8H8}, and \ch{C10H8}). Error bars shown for \ch{C6H6} represent average variation and are typical. \textbf{B:} Pressure and temperature dependence of average peak areas measured for complete range of detected PAH species.}
\label{fig:PAH}
\end{figure}

GC-MS measurements of the sampled products found a range of PAH species in the gas phase, Fig. \ref{fig:PAH}. The results are reported as averages of repeated gas sample measurements and are normalized for injection volume using the internal standard, \ch{Ar} (measured simultaneously for the same injection using GC-TCD). The adjusted values are reported as peak area and have an average variation of about 18\%; Quantitative calibration for the detected PAH species is not reported in this work. The most abundant species found were \ch{C6H6}, \ch{C7H8}, \ch{C8H8}, and \ch{C10H8} respectively which were detected in all temperature and pressure conditions, Fig. \ref{fig:PAH}\textbf{a}. Between 1 bar and 4 bar, major PAH species amounts increased significantly. The largest increase between 1 and 4 bar occurred at 1400 K. For increasing pressures above 4 bar, the PAH amounts increased but with lower sensitivity to pressure, Fig. \ref{fig:PAH}\textbf{a}. However, the sensitivity of the PAH species to temperature was found to be significant. At 1400 K the highest amounts of PAH were detected. The PAH amounts reduced sharply as the temperature increased. The high sensitivity to temperature is consistent for the four major PAH species shown, but was found to be most significant for \ch{C6H6}.

Higher temperatures were found to suppress both the amount and breadth of PAH species detected, Fig. \ref{fig:PAH}\textbf{b}. Fewer types of PAH species were detected and at lower amounts at 1800 K as compared to the 1400 K condition. Notably, the lower temperatures also found greater number of PAH with the same number of \ch{C} atoms but differing numbers of \ch{H} atoms. The reduction of PAH types and amounts at higher temperatures combined with the increase in amounts at higher pressures may indicate the dominance of chain branching PAH breakup reactions. Such reactions may be suppressed by increasing pressure and accelerated by increasing temperature, potentially in agreement with the observed trends. However, the assessment of the dominant reactions is inconclusive without more detailed PAH species measurements.

An analysis of the measured PAH trends from the simulation results found PAH mass fractions dependence on pressure shifted at different temperatures. The highest PAH concentrations were predicted at 1600 K and 1 bar and decreased with increasing pressure. At 1400 K and 1800 K, lower PAH concentrations were predicted and they increased with pressure. While the experiment trend is not replicated, a detailed analysis of the reactions can provide additional insight.

\subsection{Rate of Production Analysis}
 
\begin{figure*}[ht!]
\centering
%\vspace{-0.4 in}
\includegraphics[width=\textwidth]{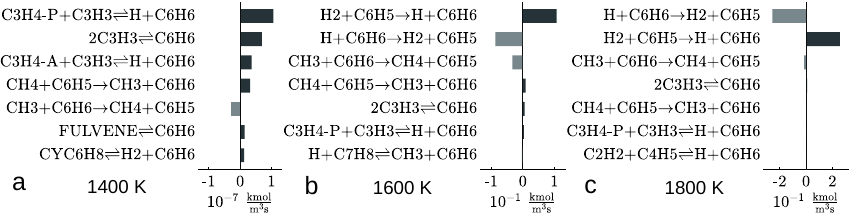}
%\vspace{4 pt}
\caption{Consumption and production rates of \ch{C6H6} given by elementary reactions in order of decreasing contribution. Analysis is based on the full CRECK mechanism (S1) in gas phase. Top 7 reactions are shown for global rate of \ch{C6H6} production from \ch{CH4} pyrolysis at 4 bar and \textbf{a:} 1400 K, \textbf{b:} 1600 K, and \textbf{c:} 1800 K.}
\label{fig:ROPAGas}
\end{figure*}

Rate of production analysis (ROPA) of the most active reactions involved in the formation and consumption of benzene (\ch{C6H6}) was performed at 1400, 1600, and 1800 K and 4 bar for the gas phase CRECK mechanism, Fig. \ref{fig:ROPAGas}. The top 7 formation or consumption reactions are reported. Benzene (\ch{C6H6}) represents the most simple PAH species containing a single aromatic ring and was found to represent the trends of larger PAH well in both experiment data and simulation predictions. The ROPA was performed for each simulation condition as a global rate analysis. The results found \ch{C6H6} formation at 1400 K was supported mainly by short chain combination reactions involving \ch{C3} species, Fig. \ref{fig:ROPAGas}\textbf{a}. At this temperature, net formation rates of \ch{C6H6} were positive and PAH concentrations were increasing. This prediction supports the low product yields predicted by the simulations, indicating the reaction is in the early stages. At 1600 K, \ch{C6H6} reactions were dominated by H-abstraction and addition reactions through the methyl and hydrogen radicals, Fig. \ref{fig:ROPAGas}\textbf{b}. The predicted reaction rates were found to be significantly higher than at 1400 K. However the net production rate of \ch{C6H6} after 60 ms was negative, indicating consumption of \ch{C6H6} in the product mixture. At the highest temperature of 1800 K, the dominant reactions were found to be the forward and reverse reactions of the H-abstraction reaction for \ch{C6H6} with the H radical ($\mathrm{H + C_6H_6 \rightarrow H_2 + C_6H_5}$), Fig. \ref{fig:ROPAGas}\textbf{c}. The forward and backward reaction rates dominated other reactions involved in \ch{C6H6} formation, indicating that the global reaction is approaching partial equilibrium for the H-abstraction reaction. The net production rate was negative, but the consumption of \ch{C6H6} at about $-5\mathrm{x}10^{-5}$ kmol/m$^3$s was lower than at the 1600 K condition by a factor of 100.

In summary, the simulations at 4 bar predict low formation rates of \ch{C6H6} at 1400 K, governed by short chain combination reactions. At 1600 K, H-abstraction reactions increased, and at 1800 K the system approached partial equilibrium for $\mathrm{H + C_6H_6 \rightarrow H_2 + C_6H_5}$ and near steady state for \ch{C6H6} production rate. Considering PAH as precursors to carbon formation, the low production rates of \ch{C6H6} at 1400 K and the approach towards partial equilibrium at 1600 K may be inhibiting the formation of sufficient concentrations of PAH to generate high carbon yields. This may partially explain the predictions of low carbon yield compared to measurements.

\begin{figure}[ht!]
\centering
\includegraphics[width=192pt]{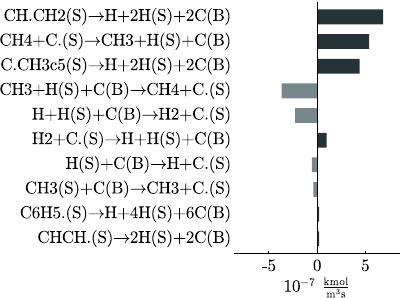}%
\vspace{4pt}%
\caption{Consumption and production rates of dominant reactions involving bulk phase carbon, C(B), in a PSR simulation performed at 10 ms of the PFR simulation using the CRECK-Surf mechanism for pyrolysis at 1600 K and 4 bar.}
\label{fig:ROPABulk}
\end{figure}

The dominant formation and consumption reactions in the CRECK-Surf mechanism were analysed for carbon deposition, Fig. \ref{fig:ROPABulk}. This analysis was performed for the global reactions involving the carbon bulk species C(B) in a single stage of the PSR simulation. The stage and condition were selected to include both high deposition rates and high PAH formation rates in the gas phase. This coincides with earlier stages of the PFR simulation with total residence times of about 10 ms. The selected condition was 1600 K and 4 bar. The dominant reactions in the surface deposition of carbon were found to be reactions involving the restructuring of the interface layer (first and third reaction in Fig. \ref{fig:ROPABulk}), which produces a hydrogen radical in the process. The second most dominant reaction was the direct deposition of a single carbon atom from the base \ch{CH4} fuel. Interestingly PAH species were not found to significantly contribute to the carbon deposition rate in a single step at the simulated conditions and instead contributed through multi step elementary reactions within surface species at the interface layer.

The findings indicate that the contribution of the heterogeneous reactions to the overall carbon yield predicted by the simulations is mainly in the creation of gas phase radicals of methyl or hydrogen. The interactions of the surface with short chain hydrocarbons or aromatics may be limited at the conditions in this study. The earlier ROPA indicated PAH mass fractions in the gas phase are dominated by elementary reactions that share the same key radicals. Deviations in the predictions of PAH reactions and their temperature and pressure sensitivity can thus lead to changes in the surface deposition mass through competition with the dominant radicals, namely methyl, short chain hydrocarbons (\ch{C2} to \ch{C3}) and H atom. It should be noted that simulations were also performed with non-constant surface site density which predicted a sharp increase in the surface site density and an over prediction of the deposition mass. While demonstrating the sensitivity of the system to site density, this may not be representative of the true site density of a graphitic carbon surface \cite{Cuoci_SurfModel}. The complexity of the deposition structure and morphology observed in several studies \cite{ThomsonDetailedMicroscopy, SVRatioItalians, MethProblems} suggests the need for more detailed treatment of the type of active sites and their surface density.
%ROPA_CB.pdf

\subsection{Assessment of the discrepancies}
\label{subsec:resultsassesment}

The simulations performed with both gas and gas/surface kinetics showed low pyrolysis yields at 1400 and 1600 K. Particularly at 1400 K, the simulations found near zero conversions of fuel to carbon. Measurements of the unreacted methane fraction found that fuel consumption is significant at 1400 K. The discrepancy can possibly be attributed to low reaction rates in the fuel decomposition reactions at low temperatures. The increase in unreacted fuel with increasing pressure at 1800 K found in both simulations and experiments was not observed in the simulations at 1400 K where the unreacted methane fraction was found to decrease with pressure. This may indicate that chain branching reactions such as $\mathrm{CH_4 \rightarrow CH_3 + H}$ which are prone to suppression at higher pressures may be under represented at 1400 K. 

Simultaneously, analysis of the PAH reaction pathways from the simulations predicts a rapid approach to partial equilibrium for dominant PAH reactions involving H-abstraction from methyl and hydrogen atoms.  At 1400 K this process was found to consume short hydrocarbon radicals in the \ch{C2} to \ch{C3} range. This may compete with radicals required for the decomposition of the fuel which may further contribute to the predictions of low fuel decomposition at 1400 K. The increase in methane decomposition reaction rates at 1600 K would provide additional radicals for the PAH growth process and may explain the higher predicted yields and mass fraction of PAH species, although still lower than the measured yields. At this intermediate temperature the simulations predict high PAH mass fractions compared to the 1400 and 1800 K conditions, while the experiment predicts high PAH mass fractions at 1400 K. Lower rates of PAH consumption reactions at 1600 K may explain the prediction of high PAH mass fractions and the low carbon and hydrogen yields compared to measurements from the global pyrolysis reaction.

The simulations with surface chemistry found that its inclusion has a minor effect on the gas composition and does not significantly contribute to the yields of pyrolysis products. This is also supported by the experiment measurements which found similar total carbon yields can occur with varying amounts of surface deposition. Analysis of the dominant reactions suggests that surface deposition rate at low residence times is dominated by availability of methyl radicals, which also directly competes with the fuel decomposition reactions. However, the simulations at 1800 K also predicted lower deposition mass than the measurements, despite the apparent availability of intermediate species and high product yields. This may indicate that the deposition reaction rates at high temperatures or the assumptions of constant surface site density could be revised. 

The mechanisms used have correctly replicated many challenging flame and flow reactor experiments used in their development \cite{Cuoci_pyrolysis_benzene, Cuoci_SurfModel}. Based on the measurements and simulations analysis of pyrolysis at the conditions in the current study they may benefit from accelerated fuel decomposition reactions at 1400 K, accelerated PAH consumption reactions at 1600 K, and increased surface deposition rates at all temperatures. Inclusion of more detailed processes such as particle diffusion, adsorption kinetics, and endothermic heat loss may also be beneficial to account for all the gas phase and surface deposition processes, further isolating the effects of the kinetics from the experiment structure and potentially addressing the discrepancies with pressure dependencies observed in conventional flames.

\section{Conclusion}
\label{sec:concl}

High temperature and high pressure methane pyrolysis reactions were studied in a micro flow reactor from 1 to 10 bar and 1400 to 1800 K. The yields of carbon in the gas phase, carbon in surface depositions on the reactor walls, hydrogen, and unreacted methane fraction were quantified. The measurements found reaction yields decreased with increasing pressure at all temperatures. This is attributed to the suppression of chain branching reactions governing the fuel decomposition.

Qualitative analysis of the detected PAH species found PAH amounts were highest at low temperatures and high pressures. Simulations using the CRECK mechanism with discrete sectional models for tracking soot formation and simulations including surface deposition mechanisms captured the pressure dependence of pyrolysis yields well at 1800 K. At 1400 and 1600 K, the simulations predicted lower yields and varying pressure dependence.

Analysis of the reaction pathways suggests the rates of fuel decomposition reactions at 1400 K and the rates of PAH consumption reactions at 1600 K may be significant to the global yield. The measured proportion of heterogeneous carbon reduced at high pressures. The surface kinetics in the simulation were not found to be a significant contributor to overall yields. Predictions of surface deposition mass were strongly linked to the gas phase concentrations of methyl and short chain radicals. Improvements to the reaction kinetics for fuel decomposition and PAH consumption at the conditions of the current study combined with the use of more detailed simulation frameworks incorporating surface morphology and diffusive effects may lead to greater insight in pyrolysis reaction processes in the micro flow reactor.

\section*{CrediT authorship contribution statement}

{\bf Mohammad Reza Razavi}: Methodology, investigation, formal analysis, visualization, writing - original draft.
{\bf \"{O}mer G\"{u}lder}: Conceptualization, resources, methodology, writing - review \& editing, supervision.

\section*{Declaration of competing interest}

The authors declare that they have no known competing financial interests or personal relationships that could have appeared to influence the work reported in this paper.

\section*{Acknowledgments}

The authors thank the Natural Sciences and Engineering Research Council of Canada for a discovery grant (RGPIN-2023-04914) supporting this research work.

%This LaTeX template was updated from previous versions for submissions to the \emph{Proceedings} and to the Symposium. It is based on the document that has been provided for Volume 42. Figure~\ref{rho_T} was taken from a LaTeX template provided by Joe Oefelein for early \emph{Proceedings} volumes. Figure~\ref{engine_fig} and Table~\ref{engine} were provided by Samuel Kazmouz from Ref.~\cite{kazmouz21}, and Table~\ref{mechanisms} is based on a table from Jun Han's Ph.D. dissertation \cite{han22}. The contributions of numerous colleagues who have contributed to this template are gratefully acknowledged by the Editors.

\FloatBarrier

\bibliographystyle{cnf-num}
\bibliography{biblio}

\end{document}

\maketitle

\section{Survey of experimental conditions}

A survey of several shock tube and flow reactor studies found the operating conditions shown in Table \ref{tab:survey}. The fuels used in the studies range from \ch{CH4} to complex fuels such as diesel. Those studies which have investigated \ch{CH4} pyrolysis were used to approximate the range of conditions for the reactor type.

	\begin{table}[!htpb]
		\caption{Survey of several shock tube and flow reactor experimental conditions.}
		\label{tab:survey}
		\begin{tabular}{cc>{\centering\arraybackslash}p{3cm}ccccc}
			\hline
			Year & Reactor & Fuel                                                & Reaction & Fuel Mole &  Pressure  & Temperature  &  Residence  \\
			     &         &                                                     &   Type   & Fraction  &   (atm)    &     (K)      &  Time (ms)  \\ \hline
			1981 \cite{evans_shock_1981} &  Shock  & \ch{C7H8}, \ch{C6H6}, \ch{n-C7H16}                  & Pyr., Ox.  &   0.01    & 0.7 to 2.8 & 700 to 2600  &      2      \\ %\hline
1995 \cite{alexiou_soot_1995} &  Shock  & \ch{C7H8}, \ch{n-C7H16}, \ch{C8H18} mixtures        & Pyr., Ox.  &   0.005   & 1.8 to 3.5 & 1500 to 2400 &     0.5     \\ %\hline
1996 \cite{alexiou_soot_1996}&  Shock  & \ch{C7H8}, \ch{CH3OH}, \ch{C2H6O}, \ch{O2} mixtures & Pyr., Ox.  &   0.01    & 1.7 to 3.8 & 1500 to 2400 &     0.5     \\ %\hline
1996 \cite{kellerer_soot_1996}&  Shock  & \ch{CH4}, \ch{C3H8}, \ch{C7H16}                     & Pyr., Ox.  &   0.04    & 15 to 100  & 1600 to 2100 &      2      \\ %\hline
2000 \cite{kellerer_measurements_2000}&  Shock  & \ch{CH4} to \ch{C7H8}                               & Pyr., Ox.  &   0.01    &  10 to 60  & 1500 to 2300 &      1      \\ %\hline
2006 \cite{sivaramakrishnan_shock-tube_2006}&  Shock  & \ch{C6H6}                                           &   Pyr.    &   0.01    &  30 to 50  & 1200 to 1800 &     1.5     \\ %\hline
2011 \cite{agafonov_soot_2011} &  Shock  & \ch{C6H6}, \ch{C7H8}, \ch{C8H10}                    &   Pyr.    &   0.01    & 0.3 to 0.5 & 1600 to 2600 &      2      \\ %\hline
2011 \cite{abanades_experimental_2011}&  Flow   & \ch{CH4}                                            &   Pyr.    &    0.5    &   1 to 1   & 1600 to 2000 & 0.5 to 40 s \\ %\hline
2012 \cite{comandini_chemistry_2012}&  Shock  & \ch{C6H5} radicals                                  &   Pyr.    &   0.01    &  25 to 60  & 900 to 1800  &      1      \\ %\hline
2012 \cite{mathieu_soot_2012}&  Shock  & Diesel                                              & Pyr., Ox.  &   0.01    &  10 to 17  & 1600 to 2500 &      3      \\ %\hline
2018 \cite{nativel_shock-tube_2019}&  Shock  & \ch{CH4}                                            &   Pyr.    &    0.1    & 1.5 to 30  & 1285 to 2400 &      5      \\ %\hline
2023 \cite{shirsath_soot_2023}&  Flow   & \ch{CH4}                                            &   Pyr.    &   0.35    &   1 to 1   & 1473 to 1873 &      5      \\ %\hline
2023 \cite{thorsen_high_2023}&  Flow   & \ch{n-C7H16}                                        &    Ox.    &  0.0005   & 20 to 100  &  400 to 900  &  6 to 50 s  \\ %\hline
2024 \cite{ferris_experimental_2024} &  Shock  & \ch{CH4}                                            &   Pyr.    &   0.35    &  1.6 to 4  & 1800 to 2400 &      5      \\ %\hline
2024 \cite{celik_role_2024} &  Flow   & \ch{CH4}                                            &   Pyr.    &   0.50    &   1 to 4   & 1300 to 1900 &  1 to 7 s   \\ %\hline
2024 \cite{punia_reduced_2024} &  Flow   & \ch{CH4}                                            &   Pyr.    &     1     &   4 to 4   & 900 to 1300  &     300     \\ %\hline
2026 \cite{clark_carbon_2026} &  Shock  & \ch{CH4}                                            &   Pyr.    &   0.05    &   4 to 5   & 1850 to 2450 &      7      \\ \hline
		\end{tabular}
	\end{table}
	
\section{Carbon burnoff measurement}
	
Details of the measurement process for solid carbon are provided. The operating principle of the carbon burnoff instrument, the procedures for sample preparation, and applications in other studies are given.

\subsection{Measurement principle}

The carbon burnoff instrument (Eltra SC800 surface carbon determinator)  operates a quartz tube furnace using pure \ch{O2} carrier gas combined with infrared absorption \ch{CO2} detectors. The samples were placed in a combustion boat and pushed into the enclosed tube furnace. The operating temperature of the furnace can be adjusted up to 1000 $^\circ$C. Once inside, surfaces of samples oxidized in the high temperature \ch{O2} environment and produced combustion products. The combustion products were filtered through \ch{H2O} traps to remove any water from the combustion process and also filtered through a packed catalyst material to convert \ch{CO} to \ch{CO2}. Finally the \ch{CO2} was passed through two infrared absorption cells of different lengths. The adsorption signal was time integrated to obtain the total \ch{CO2} mass in the products and subsequently used in the calculation of the sample's surface carbon mass. For carbon rich substrates such as steel, the combustion processes in the furnace are rate limited by \ch{O2} diffusion into the material bulk. As such they are representative of only the surface carbon. For carbon free substrates such as ceramic tubes with deposited carbon, there is no carbon products originating from the material bulk and so the surface carbon measurement is representative of the total carbon.

The instrument has high sensitivity to sample carbon mass and can detect trace carbon products as low as a few hundred ng. The instrument systems were calibrated prior to each set of measurements by injection of a controlled dose of \ch{CO2} gas into the furnace. The system was also checked periodically by combustion of carbon and sulfur in soil certified reference material (AR4015). The measurement uncertainty of the instrument is far below the uncertainty introduced due to sample contamination and typically neglected. 

\subsection{Sample preparation procedures}

After each experiment run for carbon yield measurement, gas phase carbon products were deposited on a ceramic filter at the reactor exit. A small amount of gas phase carbon deposited near the exit of the reactor wall due to thermophoresis in the cooled zone of the reactor. This was washed with deionized water and collected in a dish. The water was then absorbed onto fresh ceramic filters which are strongly hydrophilic and transfer any washed out carbon to the ceramic filter.

Surface phase depositions on the reactor wall must be prepared by cutting the reactor tube. The cuts were performed manually at 20 mm intervals using a diamond blade wet saw (about 1 mm kerf) using deionized water as a cutting fluid. After cutting each segment was rinsed in fresh deionized water to remove the contaminated water.

The cut wall segments, washed out filter, and deposition filter were placed in separate glass jars and dried in a thermal vacuum at 100 $^\circ$C. The thermal vacuum was held for about 30 min to ensure any moisture and adsorbed volatiles are removed from the carbon. Once the samples have cooled the glass jars were sealed and stored for carbon burnoff measurement. The carbon burnoff measurement process can take up to 8 hours and may span several days. The measurement order for repeated cases was periodically changed and no detectable changes in the measurements were observed, indicating samples were stable during storage. 

Repetitions of this procedure without fuel injection provided a measurement of the blank value. The cutting, handling, drying, and transfer process inevitably introduces sources of contamination to the measurements. While minimized, the trace levels of carbon measured (amounting to an average of 2.2 $\mu$g C per cut segment) are unavoidable. The combined blank value ranges between 60 to 160 $\mu$g C for all 15 segments and the ceramic filters. 

\subsection{Other applications}
Other applications of the carbon burnoff instrument in combustion are in jet fuel coking studies. For additional details of the measurement process and instrument uncertainties the reader is referred to \cite{rayneCoking}.

\FloatBarrier
\bibliographystyle{cnf-num}
\bibliography{SuppBib.bib}